\documentclass[preprint2]{aastex}
\usepackage{epsf,epsfig}

\def\la{\mathrel{\mathchoice {\vcenter{\offinterlineskip\halign{\hfil
$\displaystyle##$\hfil\cr<\cr\sim\cr}}}
{\vcenter{\offinterlineskip\halign{\hfil$\textstyle##$\hfil\cr
<\cr\sim\cr}}}
{\vcenter{\offinterlineskip\halign{\hfil$\scriptstyle##$\hfil\cr
<\cr\sim\cr}}}
{\vcenter{\offinterlineskip\halign{\hfil$\scriptscriptstyle##$\hfil\cr
<\cr\sim\cr}}}}}
\def\ga{\mathrel{\mathchoice {\vcenter{\offinterlineskip\halign{\hfil
$\displaystyle##$\hfil\cr>\cr\sim\cr}}}
{\vcenter{\offinterlineskip\halign{\hfil$\textstyle##$\hfil\cr
>\cr\sim\cr}}}
{\vcenter{\offinterlineskip\halign{\hfil$\scriptstyle##$\hfil\cr
>\cr\sim\cr}}}
{\vcenter{\offinterlineskip\halign{\hfil$\scriptscriptstyle##$\hfil\cr
>\cr\sim\cr}}}}}

\def\fdeg{\hbox{$\,.\!\!^{\circ}$}}
\def\farcm{\hbox{$\,.\!\!^{\prime}$}}

\begin{document}

\title{Polarization angular spectra of Galactic
synchrotron emission on arcminute scales}

\author{M. Tucci\altaffilmark{1,2,3}, E. Carretti\altaffilmark{4},
S. Cecchini\altaffilmark{4}, L. Nicastro\altaffilmark{5},
R. Fabbri\altaffilmark{6}, B.M. Gaensler\altaffilmark{7},
J.M. Dickey\altaffilmark{10}, N.M. McClure-Griffiths\altaffilmark{10,11}}

\altaffiltext{1}{Dipartimento di Fisica G. Occhialini, Universit\`a degli
Studi di Milano--Bicocca, Piazza della Scienza 3, I-20126 Milano, Italy.}

\altaffiltext{2}{I.N.F.N., Via Celoria 16, I-20133 Milano, Italy.}

\altaffiltext{3}{Instituto de Fisica de Cantabria, Avda. Los Castros
s/n, 39005 Santander, Spain.}

\altaffiltext{4}{I.A.S.F/C.N.R., Sezione di Bologna, Via Gobetti 101,
I-40129 Bologna, Italy.}

\altaffiltext{5}{I.A.S.F./C.N.R., Sezione di Palermo, Via Ugo la Malfa
153, I-90146 Palermo, Italy.}

\altaffiltext{6}{Dipartimento di Fisica, Universit\`a di Firenze, Via
Sansone 1, I-50019 Sesto Fiorentino, Italy.}

\altaffiltext{7}{Harvard-Smithsonian Center for Astrophysics, 
Cambridge MA 02138, U.S.A.}

\altaffiltext{10}{Bolton Fellow}

\altaffiltext{11}{Australia Telescope National Facility, CSIRO, 
PO Box 76, Epping, NSW 1710, Australia.}

\begin{abstract}

We study the angular power spectra of the polarized component of the
Galactic synchrotron emission in the 28--deg$^2$ Test
Region of the Southern Galactic Plane Survey at 1.4 GHz. These data
were obtained by the Australia Telescope Compact Array and allow us to
investigate angular power spectra down to arcminute scales. We find
that, at this frequency, the polarization spectra for $E$-- and
$B$--modes seem to be affected by Faraday rotation produced
in compact foreground screens. A different behavior is shown by the
angular spectrum of the polarized intensity $PI=\sqrt{Q^2+U^2}$. This
is well fitted by a power law ($C_{PI\ell}\propto\ell^{-\alpha_{PI}}$)
with slope $\sim1.7$, which agrees with higher frequency results and
can probably be more confidently extrapolated to the cosmological
window.

\end{abstract}


\section{Introduction and main results}

In recent years the measurement of the Cosmic Microwave
Background (CMB) polarization has become one of the major aims of a
large number of planned experiments. Its detection is however a
technological challenge: so far, we have only upper limits on the CMB
polarization level [see Staggs et al. (1999) for a review and the
recent measurements by PIQUE (Hedman et al. 2001) and POLAR (Keating
et al. 2001) experiments]. Since the CMB polarization signal is
expected to be less than 10$\%$ of the temperature anisotropies,
instrumental sensitivities of a few $\mu$K or less are required. These
will be probably reached by the forthcoming experiments: three space
missions, MAP (Wright 1999), Planck (De Zotti et al. 1999) and SPOrt
(which is completely devoted to the study of sky polarized emissions;
Carretti et al. 2002) will measure the polarization on nearly the
full sky, while several ground--based or balloon--borne experiments
are planned to observe small sky areas with high spatial resolution
[for instance, AMIBA (Kesteven et al. 2002), BOOMERanG 2K2 (Masi et
al. 2002) and BaR--SPOrt (Zannoni et al. 2002); see De Zotti (2002)
for a short review].

The possibility of extracting information on cosmological parameters from
CMB experiments is strictly related to the computation of the angular 
power spectra (APS), which in the presence of Gaussian statistics
give a complete statistical description of the CMB emission. Although
for the temperature fluctuations the APS is easily defined through the
ordinary spherical-harmonic expansion, for polarization we need two
different components, the ``electric'' ($E$) and ``magnetic'' ($B$)
modes, in order to describe the polarization intensity and orientation
(Kamionkowski, Kosowsky \& Stebbins 1997, Zaldarriaga \& Seljak
1997). In section 2.2 we will discuss the definition of the
polarization APS, with particular attention to the small scale limit. 

The detection of the CMB polarization is constrained by the presence
of foreground emissions. Different techniques have been worked out to
separate the cosmological signal from the Galactic and extra--galactic
emissions; all the methods exploit the differences in the frequency
and spatial behaviors. For this reason, the analysis of the APS has
become a common tool for studying the foregrounds: in fact, its
knowledge allows us to estimate the foreground contamination to the
CMB signal at different frequencies and angular scales (or equivalently,
spherical--harmonic index $\ell$). For the total intensity emission,
information on the spatial properties of foregrounds is limited only
to a small interval of frequencies and angular scales (see, e.g.,
Tegmark et al. 2000), and it is completely unsatisfactory between 20
and 90 GHz. For the polarization, the situation is even worse because
of the lack of high resolution surveys covering large areas.

In this paper we consider the synchrotron emission, which
is intrinsically highly polarized and is expected to be the
dominant foreground at low frequencies. Several
authors have estimated the APS of the polarized synchrotron using data 
obtained with different resolutions and from limited sky regions at
various latitudes (for a review of the up--to--date surveys, see Tucci
et al. 2000). On small sky patches, Fourier analysis has been applied
to the Stokes parameters $Q$ and $U$ (from which APS of the $E$--,
$B$--modes can be computed) and to the polarized intensity
$PI=\sqrt{Q^2+U^2}$ (Tucci et al. 2000; Tucci et al. 2001; Bruscoli et
al. 2002). Scalar and spin--weighted harmonic expansions have been
respectively used for $PI$ (Baccigalupi et al. 2001; Giardino et
al. 2002) and for $Q$ and $U$ (Giardino et al. 2002).
These computations are not equivalent in an important respect.
Fourier analysis, being adequate for small sky patches,
necessarily provides {\it local} effective spectra which for highly
non-Gaussian fields (such as the Galactic synchrotron distribution)
may be widely different from the global angular spectra. In practice
such local spectra may be more important than the global ones for the
separation of Galactic foregrounds from CMB. Further, we note that the
sum $C_{P\ell}=C_{E\ell}+C_{B\ell}$, which is the quantity usually
considered in CMB analyses, and $C_{PI\ell}$ provide different
information. This important point has been noticed in Tucci et
al. (2002), and will be discussed in Section 2.2.

Here we observe that it helps to explain some discrepancies appearing
in the literature. Estimates of the synchrotron polarization
spectra were performed from the Parkes surveys of the Southern (Duncan
et al. 1995, 1997, hereafter D97) and Northern (Duncan et al. 1999,
hereafter D99) Galactic Plane. For both surveys, sampling more than
half of the Galactic Equator, the APS are nearly independent of
longitude, and could be modeled by power laws with slopes
$\alpha_E\simeq\alpha_B=1.4\div1.5$ and $\alpha_{PI}=1.6\div1.8$ in
the $\ell$--range $100\div800$ [Tucci et al. 2000, 2001, Baccigalupi
et al. 2001, Bruscoli et al. 2002, Giardino et al. 2002 (although the
latter obtained $\alpha_{PI}=2.37\pm0.21$ in the $\ell$--range $40\div250$
from D97 data)]. Out of the Galactic Plane five patches are available
at intermediate latitudes, $5^{\circ}\le|b|\le20^{\circ}$, from the
survey by Uyaniker et al. (1999). The APS vary significantly there,
with  slopes ranging from 1 to $\sim2.5$ (Baccigalupi et al. 2001,
Bruscoli et al. 2002). At latitudes far from the Galactic plane the
only available information comes from the low-resolution survey of
Brouw \& Spoelstra (1976), covering about 40\% of the sky at five
frequencies in the range $408\div1411$ MHz. From this survey, Bruscoli
et al. (2002) found values of $\alpha_{E,B,\,PI}$ between $1\div2$ at
scales $\ell<100$, in agreement with the results from higher
resolution data. Using the same survey, Baccigalupi et al. (2001)
studied the $PI$ field and found steeper spectra, with $\alpha_{PI}
\simeq 3$.

The present paper extends the analysis of the APS of the polarized
Galactic synchrotron to arcminute scales ($\ell=10^3\div10^4$),
i.e., to angular scales smaller than in previous works by nearly one
order of magnitude. The study of the synchrotron contribution on these
scales is relevant for CMB observations. In fact, the angular scales
$300\la\ell<2000$ are expected to be those where CMB should
exhibit the highest level of polarized signal.
Moreover, at $\ell>3000$ non-linear effects on CMB
become important, producing polarized signal
stronger than the primary spectrum. These include the Vishniac,
patchy reionization and kinetic Sunyaev-Zeldovich effects
(Hu 2000, Liu et al. 2001).

We make use of high--resolution polarization data taken from a test
region for the Southern Galactic Plane Survey
(McClure--Griffiths et al. 2001, Gaensler et al. 2001, hereafter G01)
consisting of 1.4--GHz observations carried
out with the Australia Telescope Compact Array (ATCA).
We find that the $E$ and $B$ spectra can be well approximated by a
power law at $600\la\ell\la6000$, with a steep slope
$\alpha_{E,B}\simeq 2.7\div 2.9$. Moreover, we compute
the spectrum of the polarized intensity, $PI$ and find that it is
remarkably different from the above spectra, following a power law
with $\alpha_{PI}\simeq1.7$ on the whole $\ell$--range. 
We compare the ATCA APS with the APS computed in the corresponding
patch from the 2.4 GHz Parkes survey. We find a noticeable agreement
for $C_{PI,\ell}$, but not for $C_{E,B,\ell}$.
The different slopes found in the APS are interpreted in section 2.4 as
due to effects of Faraday rotation produced by foreground screens. No
evidence is found for a contribution of extragalactic point sources.

The behaviors of synchrotron APS on arcminute scales at GHz
frequencies may be interesting for information on Galactic
structure that spectra contain. They can tell us, in fact, about both
the magnetohydrodynamic turbulence in the emitting region and the
electron density fluctuations in the intervening medium (shown by
Faraday rotation). In this connection, the comparison between 1.4\,GHz and
2.4\,GHz data may be useful to separate the transverse structure of the
magnetic field in the emitting region and the longitudinal field in
the foreground screens. This point is open to future studies.

\section{Data analysis}

\subsection{The Test Region of the Southern Galactic Plane Survey
(SGPS)}

The data we use were obtained with the Australia Telescope Compact
Array (ATCA), an interferometer located near Narrabi, NSW, Australia
and consisting of five movable 22-m
antennas on a 3-km track (a sixth fixed antenna was not used in these
observations).  The data presented here were derived from observations
in multiple array configurations (see McClure-Griffiths et al. 2001),
resulting in virtually complete coverage visibilities between baselines
of 31~m and 765~m. Observations were made in the ATCA's multi-channel
continuum mode, resulting in 9 spectral channels spread across a total
bandwidth of 96~MHz, with a center frequency of 1384~MHz (see G01 for
details). There is significant Faraday rotation across the 96--MHz
bandwidth of these data, which would considerably complicate our
analysis. Therefore in the work presented here, we have only utilized
images from a single spectral channel corresponding to a center
frequency of 1404~MHz. Faraday rotation across the 8--MHz bandwidth of
this channel is negligible. The ATCA observations covered a small
portion of the sky, with Galactic coordinates
$325\fdeg5<l<332\fdeg5,\,-0\fdeg5<b<3\fdeg5$. This area
was a pilot survey for the recently completed Southern Galactic Plane 
Survey (SGPS), the analysis of which is in progress.

Maps of the Stokes parameters $Q$ and $U$ were derived from the
interferometric visibility data using the techniques described by G01. 
$Q$ and $U$ images were smoothed with a gaussian beam of FWHM
$87^{\prime\prime}\times67^{\prime\prime}$ and reprojected
into Galactic coordinates. The sensitivity of the images
was $\sim1.5\,$mJy beam$^{-1}$, except in a
strip of width 0\fdeg5 around the edges of the field for which
the sensitivity was $\sim3\,$mJy beam$^{-1}$.

Interferometric observations are sensitive only to a limited range of
spatial scales. The largest scale that can be detected is usually
determined by the shortest antenna spacing. In the ATCA case this
spacing is 31~m, which corresponds to an angular scale of
$\sim25^{\prime}$. Although the mosaicing process employed here allows
information on larger scales to be recovered (Ekers \& Rots 1979;
Cornwell 1988), it does not do so with uniform sensitivity up to the
largest scale sampled. Therefore, we here only consider angular power
spectra to be reliable at $\ell\ga600$, corresponding to an
angular scale of $\sim20$ arcmin.

The distribution of linearly polarized intensity from the Test Region
is shown in Fig. \ref{f1}. The most prominent structure is a bright
region which extends across the longitude range
$327^{\circ}<l\la331^{\circ}$, at
latitude $\sim+1^{\circ}$. A ``spur'' extends away from this region at
$l\simeq328\fdeg5$. At the edges of the survey two areas with very
low polarized emission (the ``voids'' discussed by G01)
are noticeable: the first of these is 
approximately centered in ($331\fdeg8,\,1\fdeg2$), the second
one between $327^{\circ}\la l<329\fdeg5$ at latitude
$\sim0^{\circ}$.


A wealth of small-scale structures are found in the diffuse polarized
emission. As discussed in detail by G01, some of these structures are
produced by a non-uniform distribution of magnetic fields in the
emitting regions, while in other cases small-scale structures in $Q$
and $U$ have been induced along the line of sight as a result of
Faraday rotation in compact clumps of ionized gas.  Many sets of
polarimetric observations have now identified this effect, and have
attributed  it to inhomogeneities in the interstellar medium in the
form of clouds or filaments of sizes from 1 pc to 100 pc (Wieringa et
al. 1993; Gray et al. 1999; Haverkorn, Katgert \& de Bruyn 2000).

It is interesting to compare the image in Fig. \ref{f1} with the
same area of the sky observed by the Parkes telescope at 2.4 GHz and
at the resolution of $10\farcm4$ (see Fig. 6 in G01). We expect
that polarized structures induced by Faraday rotation to be poorly
correlated between 1.4 GHz and 2.4 GHz, because
of the strong dependence of Faraday rotation and depolarization on both
frequency and angular resolution. On the other hand, we expect
a good correspondence  between the two data-sets for
polarized structures which are intrinsic to the emitting sources.
This is probably the case for the bright and extended
emission at $327^{\circ}<l\la331^{\circ}$.
Areas of fainter polarized emission seen at 1.4 GHz instead seem to be poorly
correlated with the 2.4-GHz data, indicating the effects
of foreground Faraday rotation in those regions.

\subsection{The polarization angular power spectra}

The CMB temperature anisotropies are usually expanded in spherical
harmonics, $\Delta T/T(\hat{n})=\sum_{\ell m}a_{\ell m}Y_{\ell m}
(\hat{n})$. Assuming Gaussian statistics, all information in
the CMB is contained in the angular power spectrum, defined as
$C_{\ell}=\langle| a_{\ell m}|^2\rangle$, where $\langle\,\rangle$
denotes an average over all possible realizations of the sky. The best
estimator for the APS is
\begin{equation}
C_{\ell}={1 \over 2\ell+1}\sum_m| a_{\ell m}|^2.
\label{e1}
\end{equation}
The definition of the APS for the CMB polarization is more
complex. A linearly polarized wave is described by the two Stokes
parameters $Q$ and $U$, which give us the polarized intensity,
$PI=\sqrt{Q^2+U^2}$, and the polarization direction angle
$\phi=0.5 \;{\rm atan}(U/Q)$. The value of $\phi$
depends obviously on the choice of the coordinate system; in
particular, the $Q$ and $U$ parameters change if we rotate
the coordinate system.

The standard approach to define rotationally invariant APS from $Q$
and $U$ is provided by Zaldarriaga \& Seljak (1997) (see also
Kamionkowski, Kosowsky \& Stebbins 1997 for an alternative 
approach). They describe the polarization field in terms of two
quantities scalar under rotation, $E(\hat{n})$ and $B(\hat{n})$ (in
analogy with electric and magnetic fields' properties under
parity transformation). The relation between $E$, $B$ and $Q$, $U$ is
non--local. In fact, to construct scalars under rotation at
point $\hat{n}$, we need to average the Stokes parameters around
circles centered at $\hat{n}$ (Zaldarriaga 2001). From the
harmonic expansions of the electric and magnetic modes one defines the
polarization spectra $C_{E\ell}$ and $C_{B\ell}$.

Since the analysis presented in this paper considers only small
patches of the sky, we can locally
approximate the sphere as a plane and, hence, replace the spherical
harmonic decomposition with a plane wave expansion (the standard
Fourier transforms) (Seljak 1997). This allows us to considerably
simplify the calculations. 

In the flat sky limit, we can fix a common coordinate system and use
this to define the Stokes parameters at every point $\vec{\theta}$ in
the plane of the sky. In this way, we can directly compute the Fourier
components of $Q$ and $U$,
\begin{eqnarray}
Q(\vec{\ell}) & = & \int {\rm d}\vec{\theta}
e^{-i\vec{\ell}\cdot\vec{\theta}}Q(\vec{\theta}) \nonumber \\
 & = & \int {\rm d}\vec{\theta}
e^{-i\vec{\ell}\cdot\vec{\theta}}PI(\vec{\theta})\cos(2\phi_{\vec{\theta}})
\label{e2}
\end{eqnarray}
\begin{eqnarray}
U(\vec{\ell}) & = & \int {\rm d}\vec{\theta}
e^{-i\vec{\ell}\cdot\vec{\theta}}U(\vec{\theta}) \nonumber \\
 & = & \int {\rm d}\vec{\theta}
e^{-i\vec{\ell}\cdot\vec{\theta}}PI(\vec{\theta})\sin(2\phi_{\vec{\theta}})
\label{e3}
\end{eqnarray}
where $\phi_{\vec{\theta}}$ is the angle of which the common axis must be
rotated in order to have $U=0$ in the point $\vec{\theta}$. Using the
Fourier transforms one can estimate the polarization power spectrum
through:
\begin{equation}
C_{X\ell}={1 \over 2\pi}\int
{\rm d}^2\ell^{\prime}X(\vec{\ell})X^*(\vec{\ell})\delta({\ell-\ell^{\prime}})
\label{e4}
\end{equation}
with $X=Q,\,U$.

In the small scale limit the relation between electric and magnetic
modes and the $Q$, $U$ Stokes parameters assumes a very simple
expression, consisting of a rotation in the $\ell$--space:
\begin{eqnarray}
E(\vec{\ell}) & = & Q(\vec{\ell})\cos(2\phi_{\vec{\ell}})+
U(\vec{\ell})\sin(2\phi_{\vec{\ell}}) \nonumber \\
B(\vec{\ell}) & = & -Q(\vec{\ell})\sin(2\phi_{\vec{\ell}})+
U(\vec{\ell})\cos(2\phi_{\vec{\ell}})\,.
\label{e5}
\end{eqnarray}
The $C_{E\ell}$ and $C_{B\ell}$ spectra are then computed using
eq. \ref{e4}.

The same expressions used to define the CMB polarization spectra can
be safely applied to foreground components and, in particular, to the
synchrotron emission. However, we would like to stress some
differences. First of all, the spatial distribution of synchrotron
radiation, both in the total intensity and in its polarized components,
exhibits strong non--Gaussian features and may depend on the sky
position. For instance, the Galactic plane, the North Galactic Spur and
discrete sources generally present an emission level that is much
higher than the other parts of the sky. Now, considering the Test Region, 
Fig. \ref{f1} shows, together with an extended and bright region,
areas with extremely low intensity levels. Such strong variations in
the polarization intensity are due not only to the different intrinsic
level of polarization, which we expect to be more uniform on the
Galactic plane, but also reflect the different importance of the
depolarizing mechanisms that occur inside the emitting regions or
along the line of sight. Moreover, the polarized emission shows a
weaker dependence with respect to $b$ than the total intensity.
So, the polarization APS computed from this
survey does not describe only the intrinsic spatial distribution of
polarized emission and can locally vary according to the area
considered. However, when we consider patches of the sky that are large
with respect to the dimensions of the observed structures, the angular
spectrum is expected to adequately describe the average spatial
properties of the emission. In our analysis, we consider also several
small patches ($1^{\circ}\times1^{\circ}$) in order to study
how the spectrum shape changes among regions with faint and bright
polarized emission.

Differences between CMB and foregrounds concern also
the $B$--mode. For cosmological signals induced by scalar
perturbations this quantity vanishes. It arises only in presence of
vector or tensor perturbations and in any case its intensity level is
smaller than the $E$--mode one (Zaldarriaga \& Seljak 1997). On the
contrary, the synchrotron emission contributes, on average, with the same
amount to both quantities and their spectra present a similar shape
(Seljak 1997).

In order to estimate the APS from observational data, we need to
consider in the above expression (\ref{e4}) the characteristics of the
instrument. If $X(\vec{\theta})$ is the quantity measured by ATCA 
experiment on a square grid of $N$ pixels covering a solid angle
$\Omega$, in the small scale limit the estimator of the angular power
spectrum can be computed as
\begin{equation}
C_{X\ell}={\Omega \over N_{\ell}} \sum_{\vec{\ell}}
X(\vec{\ell})X^*(\vec{\ell})b^{-2}_{\vec{\ell}}-w^{-1}\,,
\label{e6}
\end{equation}
where $X(\vec{\ell})=N^{-1}\sum
e^{-i\vec{\ell}\cdot\vec{\theta}}X(\vec{\theta})$ is the discrete 
Fourier transform of the data and corresponds to the quantities
$Q,\,U,\,E,\,B$. The sums are performed over the $N_{\ell}$
independent modes with wavevector magnitude around $\ell$. In the
present case the term $b_{\vec{\ell}}$ takes into account that the
ATCA images were subsequently smoothed with an elliptical Gaussian
window function of FWHM
$\theta_a\times\theta_b=87^{\prime\prime}\times67^{\prime\prime}$.
The expression in Fourier space is
\begin{equation}
b_{\vec{\ell}}=\exp\Bigg\{-{1 \over 2}\vec{\ell}\cdot{\bf
M}\cdot\vec{\ell}\Bigg\}\,,
\end{equation}
where the matrix ${\bf M}$ embodies the beam properties. We checked
that adopting a circular beam with $\theta_{\rm
ave}=77^{\prime\prime}$ would make very little difference. The
contribution of the noise, which is smoothed by the window function
$b_{\vec{\ell}}$, is subtracted by means of the factor
$w^{-1}=\Omega_b\sigma^2$, i.e. the pixel--independent measure of
noise ($\sigma$ is the rms noise amplitude, and
$\Omega_b=\theta_a\times\theta_b$).

Together with the electric and magnetic modes, in the following
analysis we consider the APS also for the polarized intensity,
$PI=\sqrt{Q^2+U^2}$. If $Q$ and $U$ are two Gaussian quantities, $PI$
will have a Ricean distribution and its standard deviation will be
$\sigma_{PI}=0.66\; \sigma_{Q,\,U}$. The polarized intensity is a scalar
quantity and can be expanded in ordinary spherical harmonics; its
power spectrum $C_{PI\ell}$ can be, therefore, defined in a way
equivalent to the CMB temperature fluctuations spectrum. In the small
scale limit, an expression equivalent to equation (\ref{e6}) can be
used to compute $C_{PI\ell}$. In this case, the contribution of noise
($w^{-1}_{PI}$) will be lower than those of $E$ and $B$ by a factor
$\sim0.43$. The function $b_{\vec{\ell}}$ is not exactly the antenna
beam used for $Q$ and $U$, however this can be considered a good
approximation for $b_{\vec{\ell}}$. The differences with the actual
smoothing function become meaningful only at angular scales very close
to the telescope resolution (they are responsible for the
increasing of $C_{PI\ell}$ that we find at $\ell>8000$; see figures).


The $C_{PI\ell}$ spectrum must not be confused with
the polarization power spectrum $C_{P\ell}$ defined in Seljak (1997),
where $C_{P\ell}=C_{Q\ell}+C_{U\ell}=C_{E\ell}+C_{B\ell}$, and clearly
has a different physical meaning with respect to $C_{PI\ell}$. The
latter, unlike the $E$-- and $B$--modes, does not provide a complete
description of the polarization field because it is related only to
the intensity of the polarization without giving any information on
its position angle. Only if the polarization angle is uniform
inside all the survey area, the equality $C_{PI\ell}=C_{P\ell}$
is ensured, while in general $C_{PI\ell}$ and $C_{P\ell}$ are
expected to show different behaviors. This is surely the case for CMB
polarization, as shown in Fig. \ref{f2}. There we report the
$E$--mode (solid line) and $PI$ (dashed line) spectra resulting from
Fourier analysis on CMB simulated maps. We chose a standard
cosmological model with only scalar perturbations, where therefore
$C_{E\ell}=C_{P\ell}$, and simulation boxes of
$10^{\circ}\times10^{\circ}$.

As discussed in the Section 1, according to previous analysis
of the synchrotron maps by Bruscoli et al. (2002), the polarized
intensity APS has a slightly faster decrease than the $E$ and
$B$--mode APS at $\ell\le800$. This is not surprising because we
expect that the $E$ and $B$ modes vary more rapidly than $PI$, due to
changes both in the intensity and in the angle of polarization.

\subsection{Results}

We study the APS of the synchrotron
polarization in several square patches of the survey with 
different dimensions.
Because of the limited sky area considered, the
Fourier approach is suitable; after computing the Fourier components
of the data, $Q(\vec{\ell})$, $U(\vec{\ell})$ and $PI(\vec{\ell})$,
equation \ref{e6} is applied to estimate the spectra for the modes
$E$, $B$ and $PI$ (the spectra for $Q$ and $U$ are redundant, as shown
in the previous section). A standard least--squares method was used
to fit each curve to a power law, $C_{X\ell}=A_X\ell^{-\alpha_X}$, in
suitable $\ell$--ranges (see Tucci et al. 2000).

{\begin{table*}[h]
\caption{Best--fit parameters for angular power spectra}
\begin{tabular}{ccccccccc}
\hline
 Survey$^\dag$ & Box & Centre & $A_E$ & $\alpha_E$ &
$\alpha_B$
& $A_{PI}$ &
 $\alpha_{PI}$ & $\ell$--range \\
  & (deg) & (l,b) & (K$^2$) & & & (K$^2$) & &
($\times10^3$) \\ 
\hline
 D97 & $5\times5$ & (329,1.5) & 9.1$\times10^{-4}$ &
 1.78$\,\pm$0.18 & 1.62$\,\pm$0.19 & 3.6$\times10^{-4}$ &
 1.68$\,\pm$0.30 & (0.1,0.8) \\
\hline
 G01 & $4\times4$ & (329,1.5) & 93.1 & 2.85$\,\pm$0.07 &
 2.74$\,\pm0.06$ & 4.2$\times10^{-3}$ & 1.66$\,\pm$0.07 & (0.6,6) \\
& $3\times3$ & (330.5,1.5) & 60.2 & 2.76$\,\pm$0.09 &
 2.65$\,\pm0.10$ & 6.0$\times10^{-3}$ & 1.67$\,\pm$0.08 & (0.6,6) \\
& $3\times3$ & (327.5,1.5) & 63.0 & 2.84$\,\pm$0.09 &
 2.70$\,\pm0.10$ & 3.7$\times10^{-3}$ & 1.68$\,\pm$0.08 & (0.6,6) \\
\hline
\end{tabular}

$^\dag$ For the D97 survey the $A_X$ values are at 2.4 GHz, for the
Test Region (G01) at 1.4 GHz. In the last column we report the range
of angular scales considered in the fit.
\label{t1}
\end{table*}


Fig. \ref{f1} shows the areas of the survey where the power spectra
are computed: a $4^{\circ}\times4^{\circ}$ box centered in 
$(l,\,b)=(329^{\circ},1\fdeg5)$, which is the largest square area
that can be extracted from the survey; two $3^{\circ}\times3^{\circ}$
boxes covering nearly all the high--sensitivity part of survey; and
six $1^{\circ}\times1^{\circ}$ boxes, in order to sample regions with
different features in polarization. 

In the Test Region, 21 compact sources showing linear polarization
were identified (see Table 1 in G01).
Some of these can be readily seen in Fig. \ref{f1}. We tried
to choose the $1^{\circ}\times1^{\circ}$ boxes in regions where the
number of compact sources is not more than one. Also for
the larger boxes we can fairly suppose that the number of compact
sources is low enough not to affect the estimate of the power
spectra.

First of all, let us discuss the results for $C_{E\ell}$ and
$C_{B\ell}$ in the $4^{\circ}\times4^{\circ}$ area. In Fig. \ref{f3}
we report only $C_{E\ell}$ (solid line); the results for $B$--mode are
similar to those for $E$--mode. In the range $600<\ell\le6000$ such
spectra are well approximated by a power law with
$\alpha_E\simeq\alpha_B\simeq2.7\div2.9$ (see Table \ref{t1}).
We compare the $E$--mode spectrum to that
resulting from the observations in the same sky area by the Parkes
telescope (D97). In the plot the Parkes spectrum has been scaled to
the ATCA frequency, assuming the frequency spectrum of synchrotron
emission $T_{\rm syn}\propto\nu^{\beta}$ with spectral index $\beta=-2.5$
and $-3$ (dashed lines). We compute the best fit for
$100\le\ell\le800$ and find a power law index of
$\alpha_E=1.78\pm0.18$. This value is significantly
lower than the best--fit $\alpha_E$ found from the ATCA survey, i.e.
$2.85\pm0.07$ (dotted line). Moreover, the amplitudes of the Parkes and
ATCA spectra show a difference of nearly one order of magnitude at
scales $600\la\ell\la1000$. In this range,
corresponding to angular dimensions between $20^{\prime}$ and
$10^{\prime}$, both Parkes and ATCA are able to detect polarization
signals. This gap, therefore, must be related to the different
frequency at which the observations were made and cannot be attributed
to our choice for the synchrotron spectral index. In fact, to remove
the discrepancy one requires a spectral index of $-5\div-4$, i.e.,
values too far from the average estimate between these frequencies
($\beta=-2.8$, see Platania et al. 1998), even in a very peculiar
region.

Quite different results come from the analysis of the polarized
intensity spectrum. For the same $4^{\circ}\times4^{\circ}$ survey patch
(see Fig. \ref{f4}) the $C_{PI\ell}$ is well approximated
by a power law with $\alpha_{PI}=1.66\pm0.07$ (the dotted line is the
best--fit in $600<\ell\le6000$). The $PI$ spectrum shows a much
flatter shape than $C_{E,B\ell}$, and looks like the
extension at high $\ell$ of the curves obtained from 2.4\,GHz Parkes
data.

Similar results are obtained from the analysis of smaller areas. In Fig.
\ref{f5} we directly compare the electric and magnetic spectra with
$C_{PI\ell}$, computed on two $3^{\circ}\times3^{\circ}$ regions that
sample all the low--noise part of the survey. Table \ref{t1} reports
the best--fit parameters of the curves. These plots make evident how
the $E$-- and $B$--mode spectra have a similar shape to each other but
separate from $C_{PI\ell}$ as they move toward low $\ell$.

{\begin{table*}[h]
\caption{The best--fit parameters for 
$1^\circ\times1^\circ$ areas in the $\ell$--range $[1000,\,6000]$}

\begin{tabular}{cccccccc}
\hline
Box Label$^\dag$ & Box Centre & $PI_{\rm rms}$ & $A_E$ &
$\alpha_E$ &
$\alpha_B$ & $A_{PI}$ & $\alpha_{PI}$ \\ 
 & ($l$, $b$) & (K) & (K$^2$) & & & (K$^2$) & \\ 
\hline
a & (330.6,1.0) & 0.46 & 168 & 2.84$\,\pm$0.25 &
2.58$\,\pm$0.24 &
2.1$\times10^{-2}$ & 1.79$\,\pm$0.20 \\
b & (329.7,1.1) & 0.54 & 2.9$\times10^3$ & 3.18$\,\pm$0.26 &
3.24$\,\pm$0.30 & 0.89 & 2.24$\,\pm$0.22 \\
c & (327.7,1.1) & 0.38 & 137 & 2.91$\,\pm$0.30 &
2.97$\,\pm$0.28 &
3.4$\times10^{-2}$ & 1.93$\,\pm$0.27 \\
d & (328.5,2.6) & 0.38 & 1.8$\times10^5$ & 3.84$\,\pm$0.34 &
3.53$\,\pm$0.35 & 5.1 & 2.58$\,\pm$0.29 \\
e & (327.0,2.3) & 0.26 & 616 & 3.18$\,\pm$0.35 &
3.07$\,\pm$0.32 &
1.0$\times10^{-2}$ & 1.88$\,\pm$0.38 \\
f & (330.4,2.3) & 0.31 & 233 & 3.01$\,\pm$0.38 &
2.64$\,\pm$0.39 &
4.4$\times10^{-3}$ & 1.68$\,\pm$0.35 \\
\hline
\end{tabular}

$^\dag\,$The letters correspond to the panels in Fig. \ref{f6}.
\label{t2}
\end{table*}

In Fig. \ref{f6} we plot $C_{E\ell}$ and $C_{PI\ell}$ obtained in
$1^{\circ}\times1^{\circ}$ regions, chosen between high-- and
low--polarized emission areas (see the rms values of
the polarized intensity $PI_{\rm rms}$ in Table \ref{t2}). We use
letters $a$ through $f$ to label six areas. The best--fit
parameters for the curves are reported in Table \ref{t2}.
We found that the slope of both $E$--, $B$--mode and $PI$ spectra in
$1^{\circ}\times1^{\circ}$ regions are quite independent of the area
considered, with only small fluctuations around the values found in
the bigger regions. For example, the spectral slope does not
significantly differ between the regions $a$ -- $b$, that are within
the brightest area of the Test Region, and $e$ -- $f$, where only
faint polarized emission is observed. The only exception is the $d$
case, which corresponds to a peculiar region inside which 1.4 and
2.4\,GHz data show anti--correlation, probably produced by internal
Faraday depolarization occurring at 1.4\,GHz. In this area
$C_{E,B\ell}$ and $C_{PI\ell}$ are much steeper than in the
other regions, indicating that internal depolarization erases
intrinsic small--scale structures.


\subsection{ISM structures, Faraday screens and point sources}

The strong differences between $C_{PI\ell}$ and $C_{E,B\ell}$ found at
1.4 GHz but not at 2.4 GHz, as well as the amplitude gap in $E$--mode
APS shown in Fig. \ref{f3} can be interpreted if
we assume that polarization structures are not intrinsic to the
emitting regions. As already discussed by G01 and
in section 2.1, an alternative mechanism to produce small--scale
structures is the Faraday rotation along the line of sight due to
foreground screens. In detail, if a smooth polarized background
emission passes through an interstellar medium (ISM) with spatial
variations in the rotation measure, the linearly polarized
radiation is affected by a Faraday rotation of the polarization
angle that varies spatially with the line of sight. Hence, if
the angular dimension of these ISM structures are sampled by the
interferometer, the Faraday rotation will induce
detectable variations in $Q$ and $U$, and the $E$ and $B$
spectra will gain extra power on these scales. Visual inspection of
Fig. 5 of G01, reporting the
polarization angle, apparently shows many sharp-edge structures with
size of the order of $10^{\prime}$, supporting our interpretation.

The situation is quite different for the polarized intensity spectrum,
$C_{PI\ell}$. The $PI$ value is not changed by a rotation of the
polarization angle, but it could be affected by depolarization effects
that the rotation induces. Some areas, where Faraday rotation may be the
cause of a strong reduction of the polarized intensity, are found by
G01: for instance, the existence of two large regions devoid of
polarized emission is
explained by beam depolarization due to large RM variations on small
scales. Their position is, however, at the edges of the survey and
their contribution to the APS of our selected regions can be
considered negligible.
Internal Faraday depolarization could be the cause
of the anti--correlation of the polarized emission 
between 1.4--GHz and 2.4--GHz data, observed in the ``spur''
region (labeled as $d$, cfr. Section 2.3). However, apart from this
peculiar region, the low RM values
measured in the rest of the ATCA survey suggest that Faraday
depolarization should not affect our estimates of the polarized
intensity spectrum, especially in the large boxes. This is confirmed
by $\alpha_{PI}$ values found in $1^{\circ}\times1^{\circ}$ regions,
which result weakly dependent on the area considered and on the
magnitude of RM measured there (see Table \ref{t2}). So, in the
absence of depolarization, the Faraday rotation changes only the
polarization angle without affecting the intensity and consequently
$C_{PI\ell}$.

The effects of Faraday screens on the polarization emission have been
discussed by previous authors.
Wieringa et al. (1993) were the first to notice that the polarized
component of the galactic background at 325 MHz is characterized by a
patchy pattern on arcminute scales. They detected features like narrow
filaments of $5^{\prime}$--$10^{\prime}$ width and ``clouds'' with
angular sizes of some tens of arcminutes, across which there are only
small changes in the polarization angle. These structures were 
found at different latitudes and were just interpreted in terms of
Faraday modulation by foreground ionized gas clumps. More recent
interferometric observations (Duncan et al. 1998; Gray et al. 1999;
Haverkorn et al. 2000) confirmed these features in polarization
even at 1.4 GHz.

In Faraday--induced polarization structures observed by Wieringa et
al. (1993), excess RMs from compact screens were measured with typical
magnitude $\sim5$ rad$\,$m$^{-2}$, corresponding to $\Delta\phi\sim
13^{\circ}$ at 1.4 GHz. Although small, these RM values suffice
to produce structures which are detectable by the ATCA.
Following the appendix in G01, if uniformly polarized emission with
intensity $P_0$ passes through a compact cloud of rotation measure
$R_c$, the interferometer detects polarized emission in the
direction of the cloud with apparent intensity
\begin{equation}
P_{\rm det}=2P_0\sin(R_c\lambda^2)\,,
\end{equation}
and measures the RM value of the cloud as $R_c/2$. Then,
using $R_c=10$ rad$\,$m$^{-2}$, at 1.4~GHz, the polarization
induced by a compact cloud will be close to 100\% of the background
component. At 2.4 GHz (the frequency of the D97 survey) this
percentage is reduced by about a factor of three, without considering
the depolarization effects due to the lower resolution.

The above discussion assumes that none of the features of angular 
spectra should be attributed to extragalactic point sources. 
This is consistent with estimates: from Tegmark and Efstathiou (1996), 
using a VLA sample of 1.5 GHz sources, flat {\it intensity} 
spectra $C_{I\ell }^{\rm {PS}}=(3\div 13)\times 10^{-8}$ K$^2$ are 
derived for limiting fluxes in the range $(0.1\div 1)$ Jy. From Toffolatti 
et al. (1998) considering an evolution model for galaxies, for the same 
limiting fluxes and the (frequency) spectral index in the range 
$-(2 \div 2.3)$ we get $C_{I\ell }^{\rm {PS}}=(0.2\div 6.6)\times 
10^{-8}$ K$^2.$ A more conservative upper limit can be derived from 
our own analysis of intensity spectra in U99 (Tucci et al. 2001,
Bruscoli et al. 2002). 
Some patches exhibit very flat spectra there, $\alpha _I\simeq 0;$ 
assuming that they are dominated from point sources, we get the limit 
$C_{I\ell }^{\rm {PS}}<5\times 10^{-7}$ K$^2.$ From this number, 
adopting a radio-source polarization degree of 5\% [in agreement with 
De Zotti et al. (2000)] we get $C_{X\ell }^{\rm {PS}}<1.3\times 
10^{-9}$ K$^2$ for $X=P$, $PI,$ and we conclude that the contribution 
of point sources should be negligible in the whole range $\ell\le6000$.
This is certainly consistent with the simple behavior
of the angular spectra that we find.

\section{Discussion}

In this paper, for the first time, we extend the study of the angular
power spectrum for the polarized
component of the Galactic synchrotron emission to arcminute scales, i.e. up
to $\ell\sim10^4$. To reach such scales we needed high--resolution
data, which were provided by the ATCA observations
of a small patch of the Galactic Plane at 1.4 GHz.

In the paper we compute the polarization spectra for ``electric'' and
``magnetic'' modes, plus the spectrum for the polarized intensity. We
find that, in the range $600\le\ell\le6000$, both $C_{E\ell}$ and
$C_{B\ell}$ can be well approximated by power laws with slopes
$\alpha_E\simeq\alpha_B\sim 2.7\div 2.9$. Such spectra are
significantly steeper than those arising at $\ell\le800$ from
low--resolution data. Moreover, their amplitude, if compared to the
spectra obtained by the Parkes telescope in the same sky area, turn
out to be
higher by nearly one order of magnitude at angular scales between
$20^{\prime}$ and 10$^{\prime}$. These peculiar behaviors are well
interpreted as due to the small--scale modulation of a relatively
uniform polarized background by Faraday rotation along the line of
sight. On the contrary, we believe that our estimates of $C_{PI\ell}$,
whose slope ($\alpha_{PI}\sim1.7$) is in agreement with D97 data at
2.4 GHz, are not affected by Faraday effects and fairly describe the
intrinsic spatial distribution of the polarized emission.

An interesting point, which arises from our analysis, regards the
distinctive meaning of $C_{PI\ell}$ with respect to $C_{E,B\ell}$.
As we have discussed in section 2.2, $PI$ is a scalar quantity and
refers only to the intensity of the polarization without any
information on its direction. We then expect that the APS
for $E$-- and $B$--modes, that provide a complete description of the
polarization field, do not have the same shape as the $PI$ spectrum. The
differences should be greater when the direction of polarization
changes very rapidly. Deviations
between $C_{E\ell}$ and $C_{PI\ell}$ are found in the CMB (see the
results of the simulations in Fig. \ref{f2}): these are not unexpected,
because of the geometry of the polarization angle in the $E$--mode
spots. In the case of synchrotron emission, the polarization
direction for the diffuse component is quite smooth on large
scales following the Galactic magnetic field. From low--resolution
surveys the estimates of $C_{E,B\ell}$ and $C_{PI\ell}$ give only
moderate differences in the spectral shape (see Bruscoli et al. 2002).
However, when small scales are considered, fluctuations in magnetic
fields, discrete sources and also Faraday effects contribute to
amplify the variations in the polarization spectra, as the present
results highlight.

The extrapolation of our results for $C_{E\ell }$ and $C_{B\ell }$
to higher frequencies should not be regarded as a trivial matter,
since Faraday effects are substantial at 1.4 GHz while they
become negligible at a few tens of GHz. The electric and magnetic
spectra are strongly affected by the Faraday rotation along the line
of sight, showing a steep slope ($\alpha_{E,B}\sim2.8$). The polarized
intensity spectrum, instead, can be more reliably extrapolated to the
``cosmological'' frequencies, because, as discussed in section 2.4, it
is not affected by Faraday rotation, except in severe cases when
significant depolarization is occurring. The power index $\alpha_{PI}$
is less than 2 independent of the region analysed, with a
value of $1.66\pm0.05$ in the $4^{\circ}\times4^{\circ}$ box. 

We have seen that analyses on the synchrotron polarization spectrum
in the literature indicate a moderate slope ($\alpha_{X}\la2$ with
$X=E,B,\,PI$) on angular scales $\ell<10^3$. Now ruling
out a significant contribution from point sources, we confirm this
result also at $\ell\la10^4$ for $C_{PI\ell}$, and we show that
synchrotron emission is rather rich in
small scale structures. Hence, contrary to what is usually assumed, it
might be a relevant contaminant in CMB polarization measurements
at very small scales. It remains to be seen if the APS as deduced
by ATCA observations of the Galactic plane is a common feature in
regions at high galactic latitudes. In general we expect the polarized
synchrotron emission to be fainter in regions far out of the
Galactic plane, except in very bright areas. In one such region
Bruscoli et al. (2002) estimate $C_{X\ell}$ for $\ell<100$,
finding a spectrum behavior consistent with those of the Galactic
plane ($1<\alpha_{X}<2$). Even if very bright regions are not
typical at high latitudes, these play an important role in the process
of foreground subtraction in CMB experiments; this is why we have to
put great care into the study of the APS of these regions.

\acknowledgments

We thank the whole SPOrt collaboration team and, in particular,
M.T. wishes to thank S. Bonometto for encouragement and useful
discussions. This work is supported by the Italian Space Agency (ASI).
M.T. acknowledges the financial support provided through the European
Community's Human Potential Programme under contract HPRN-CT-2000-00124,
CMBNET. The Australia Telescope is funded by the Commonwealth of
Australia for operation as a National Facility managed by CSIRO. B.M.G.
acknowledges the support of a Clay Fellowship awarded by
Harvard--Smithsonian Center for Astrophysics. N.M.Mc-G. and
J.M.D.  acknowledge the support of NSF grant AST-9732695 to the
University of Minnesota.

\figcaption{Linearly polarized intensity, $PI=\sqrt{Q^2+U^2}$, in the
Test Region. The image is obtained combining all nine spectral
channels. The intensity scale runs from 0.4 to 9.5 mJy
beam$^{-1}$. The square boxes indicate the areas where we computed the
angular power spectra.
\label{f1}}

{\begin{figure*}[htb]
\centerline{\epsfxsize=16truecm
\epsfbox{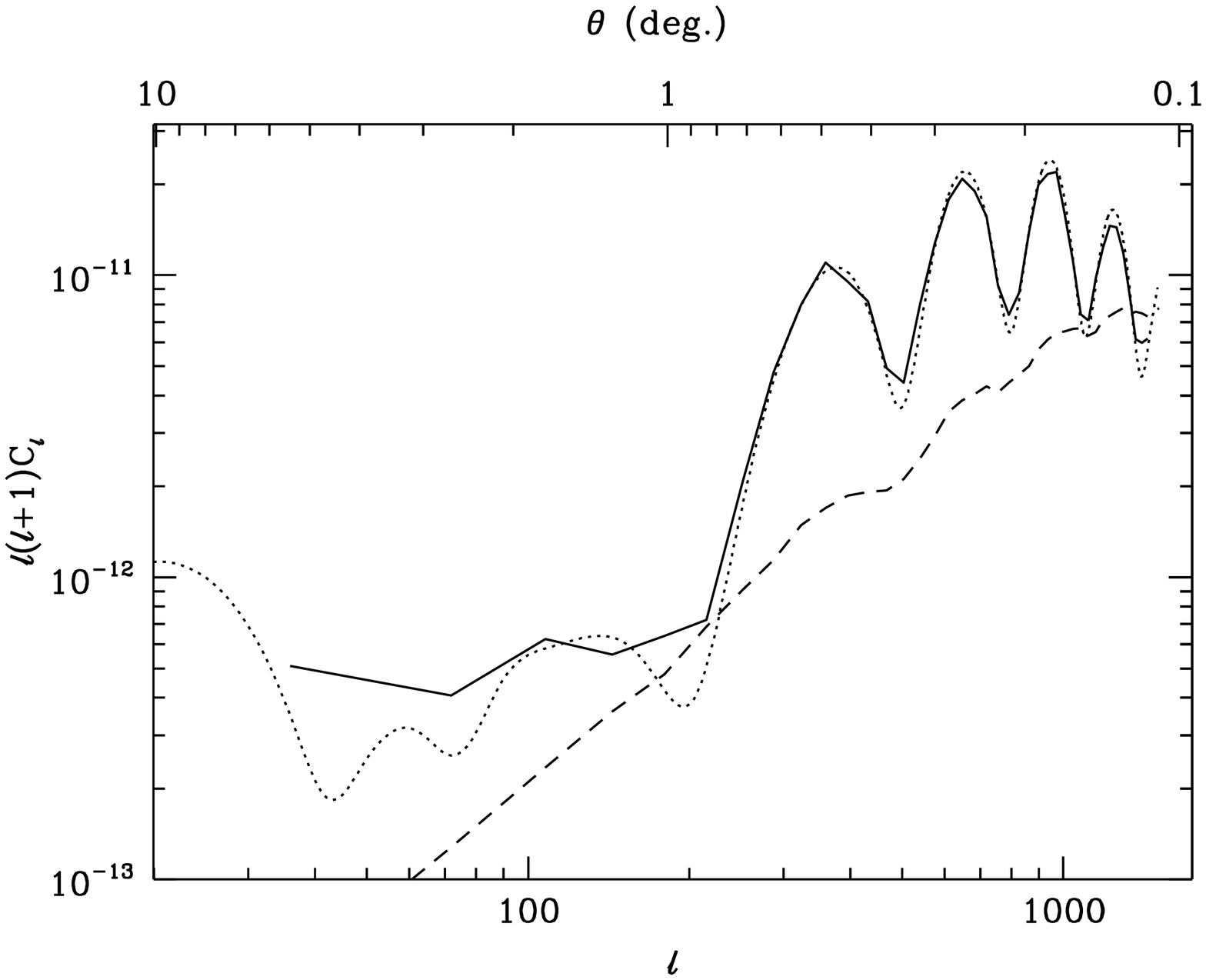}}
\caption{The CMB spectra for $E$--mode (dotted line is the input
spectrum, the solid line is the computed spectrum from Fourier
analysis) and for the $PI$ field (dashed line), from a CDM model
with a secondary ionization optical depth $\tau_{ion}=0.2$.
\label{f2}}
\end{figure*}}

{\begin{figure*}[htb]
\centerline{\epsfxsize=16truecm
\epsfbox{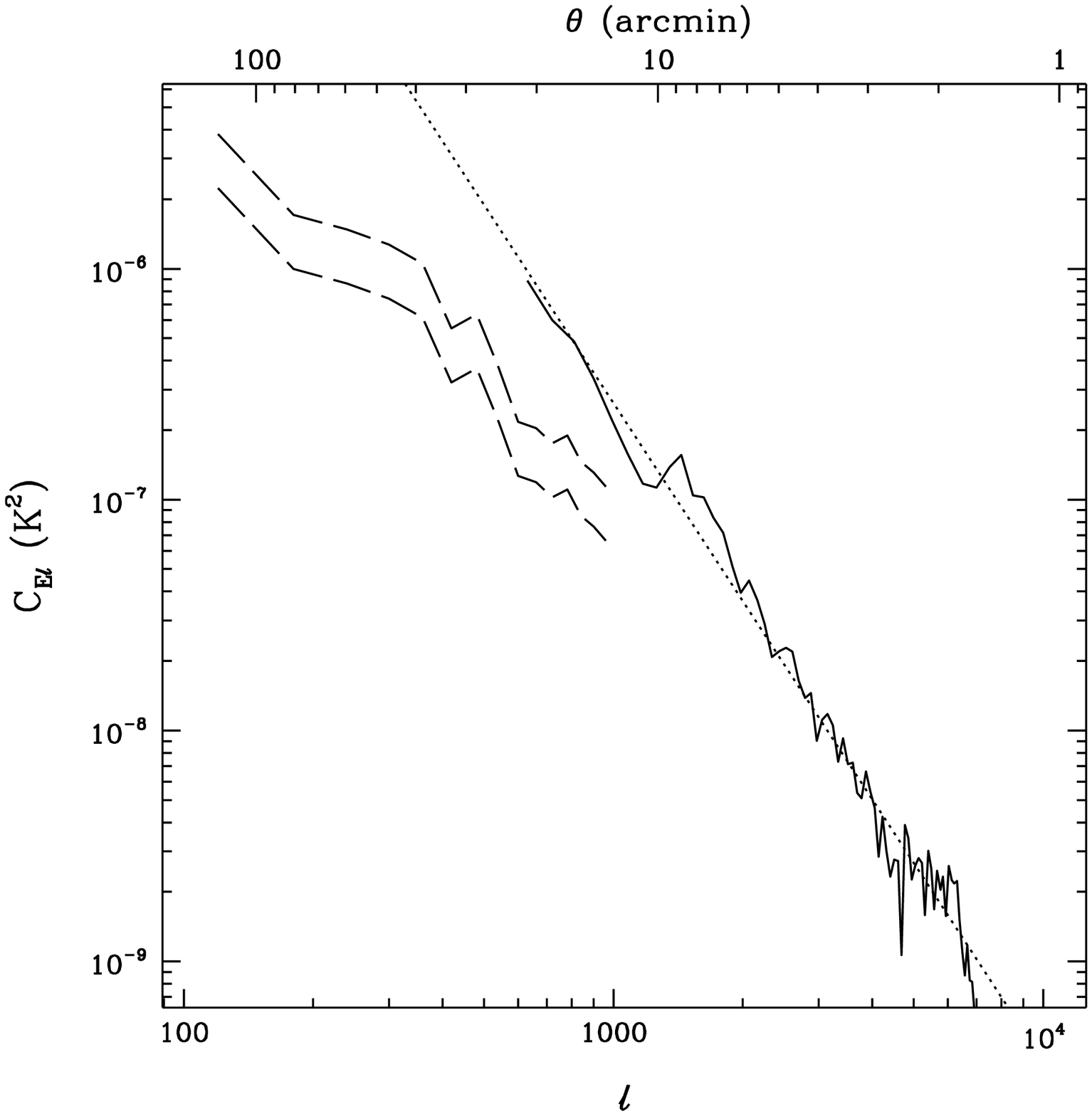}}
\caption{E--mode spectrum from the $4^{\circ}\times4^{\circ}$
region (solid lines), compared to the result from D97 data (dashed
lines) in a $5^{\circ}\times5^{\circ}$ area centered on the same
position. The dotted line is the best--fit power law in the
$\ell$--range between 600 and 6000 (see Table 1).
\label{f3}}
\end{figure*}}

{\begin{figure*}[htb]
\centerline{\epsfxsize=16truecm
\epsfbox{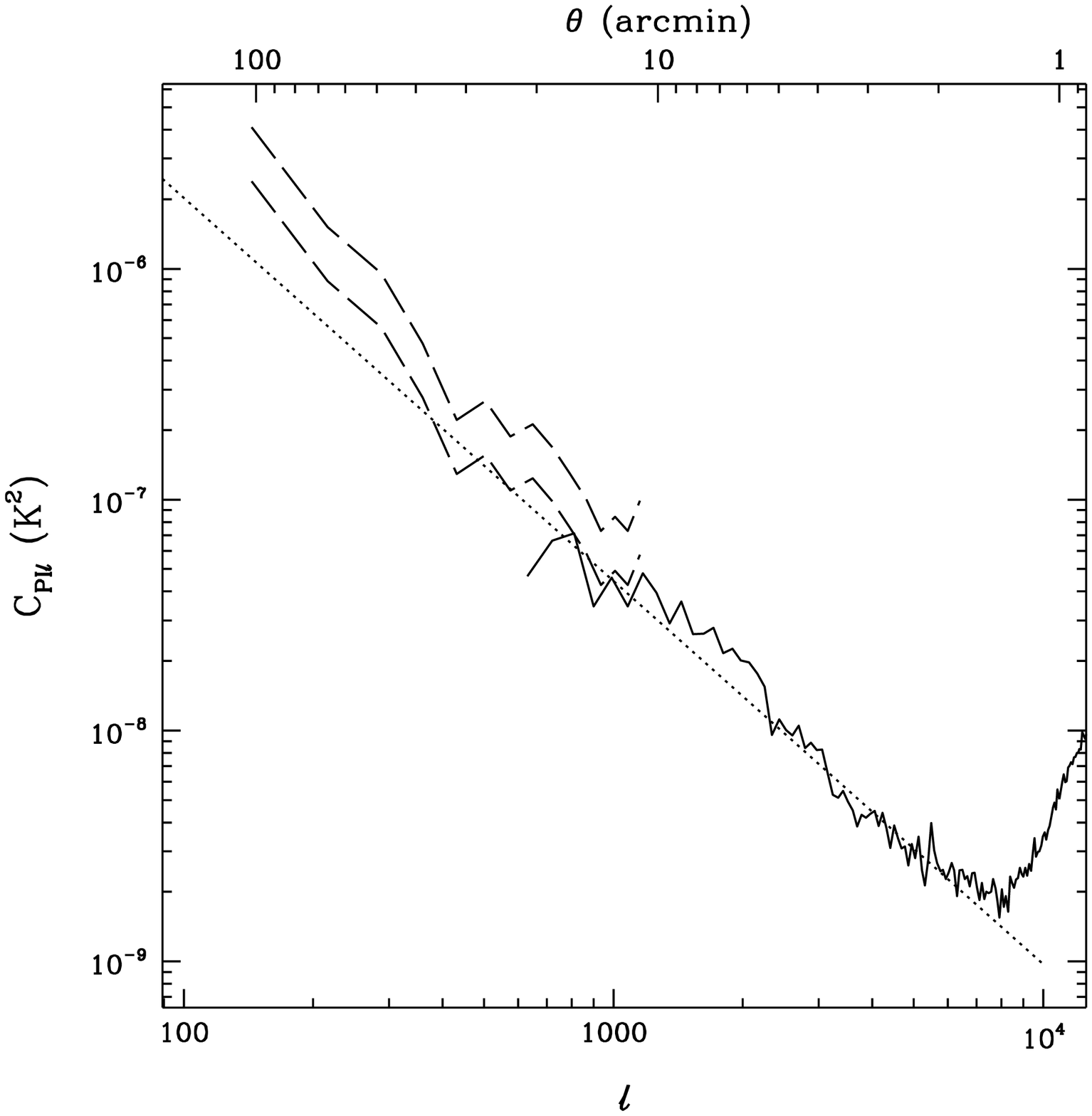}}
\caption{As in Fig. \ref{f3}, but for polarization intensity spectra.
\label{f4}}
\end{figure*}}

{\begin{figure*}[htb]
\centerline{\epsfxsize=18truecm
\epsfbox{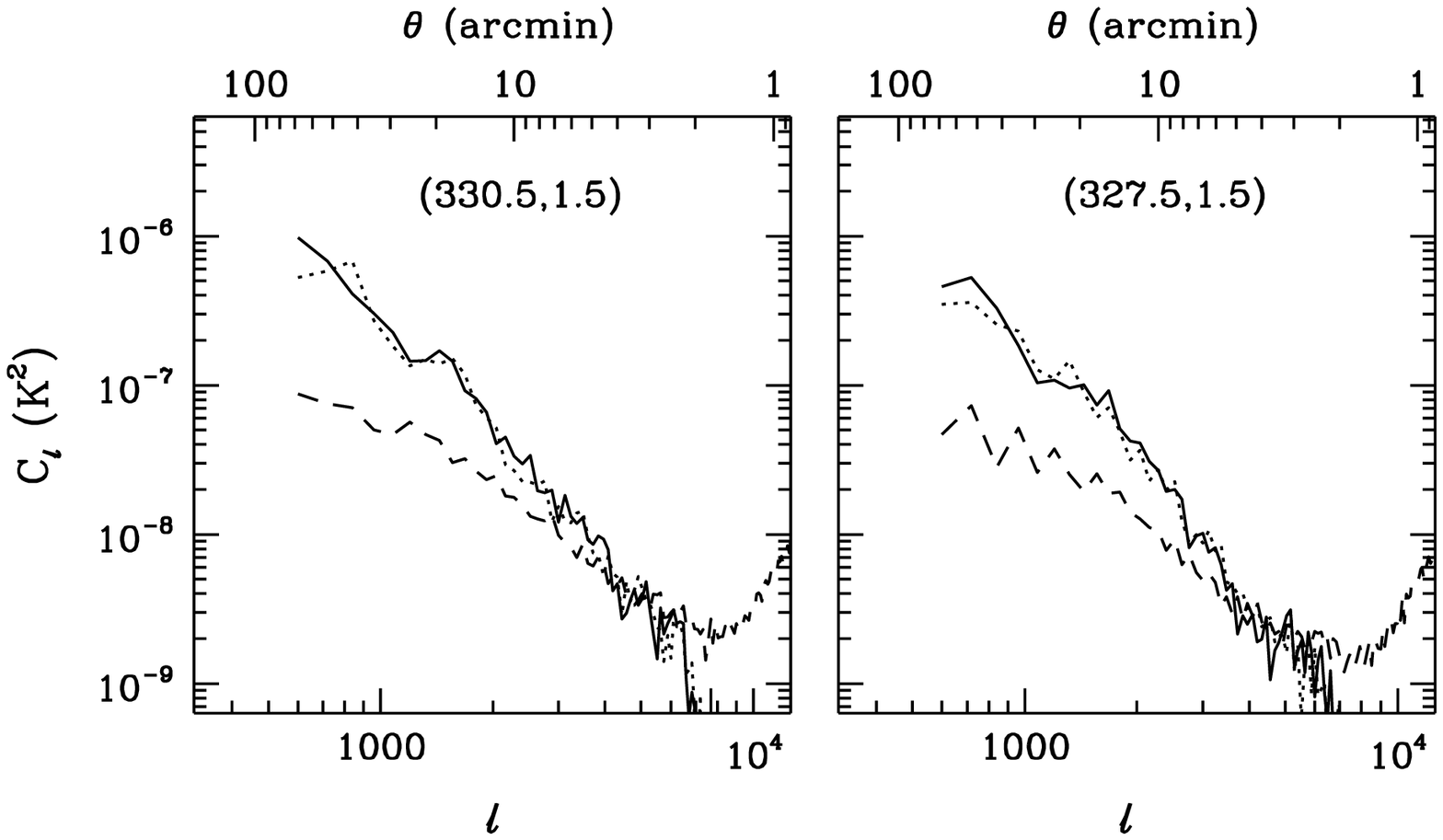}}
\caption{$C_{E\ell}$ (solid lines), $C_{B\ell}$ (dotted lines) and
$C_{PI\ell}$ (dashed lines) in two $3^{\circ}\times3^{\circ}$
boxes, whose centers are shown in the panels.
\label{f5}}
\end{figure*}}

{\begin{figure*}[htb]
\centerline{\epsfxsize=18truecm
\epsfbox{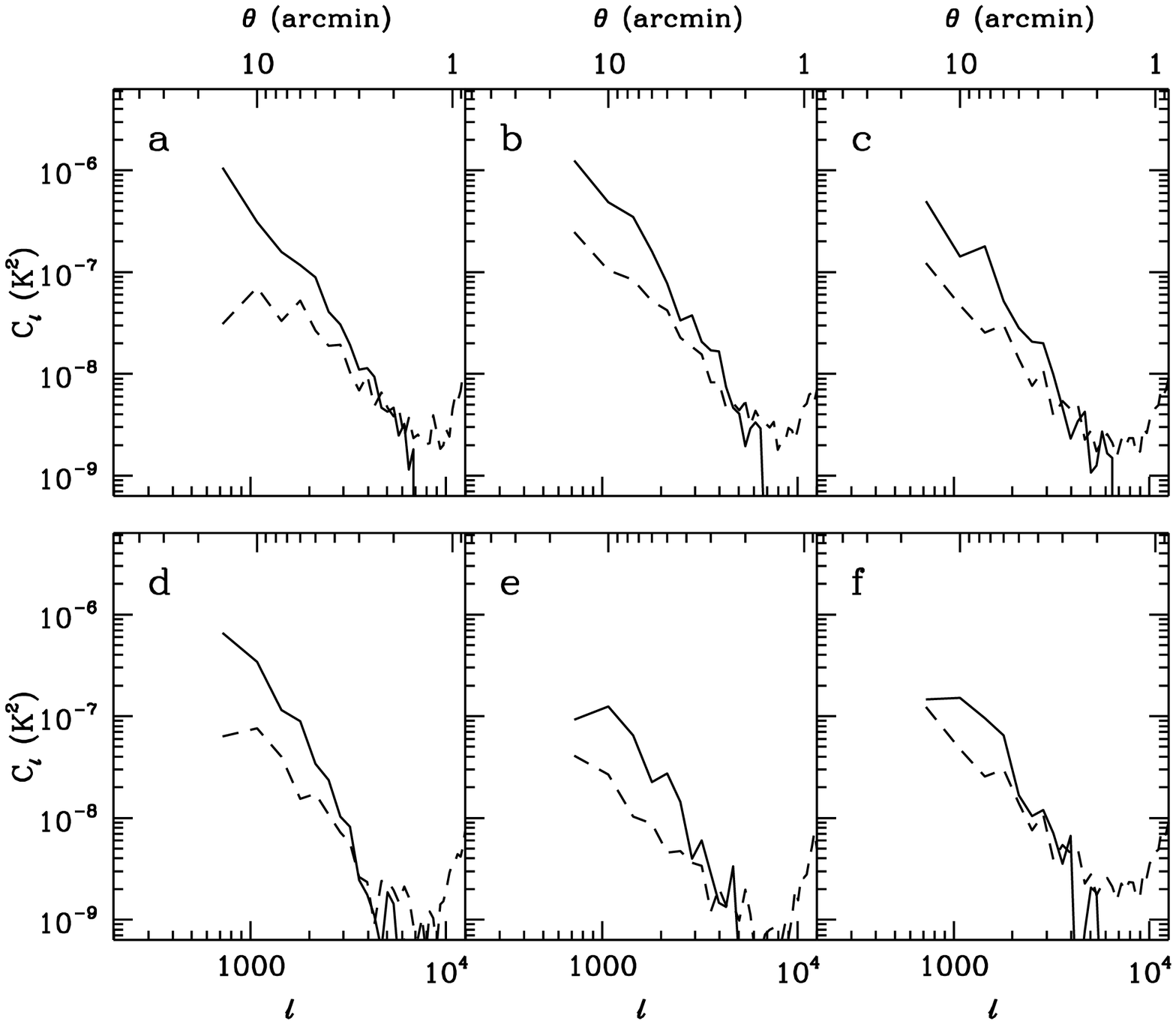}}
\caption{$C_{E\ell}$ (solid lines) and $C_{PI\ell}$ (dashed lines) in
six $1^{\circ}\times1^{\circ}$ boxes. The labels in the plots identify
the regions where the spectra are computed.
\label{f6}}
\end{figure*}}

\end{document}